# Universal scaling of *c*-axis dc conductivity for the underdoped high-temperature cuprate superconductors

T. Honma[1] and P. H. Hor[2]

[1]*Department of Physics, Asahikawa Medical College, Asahikawa, Hokkaido 078-8510, Japan Asahikawa*

[2]*Department of Physics and Texas Center for Superconductivity, University of Houston, Houston, Texas 77204-5005, USA*

Coexistence of the “metallic-like” *in-plane* and the “semiconducting-like” *out-of-plane* (*c*-axis) dc conductivities ($\sigma_c$), generating a huge anisotropy in the underdoped high-temperature cuprate superconductors (HTCS), defies our current understanding of metal. In this report we present an intrinsic doping dependence of $\sigma_c$. We find that the $\sigma_c$ for the underdoped HTCS is universally scaled to the $\sigma_c$ at the optimal doped-hole concentration. The universal scaling behavior suggests that there are three intrinsic processes contribute to $\sigma_c$: (i) the doping-dependent-activated gap; (ii) the exponential doping dependences and (iii) the tunneling between adjacent $CuO_2$ block layers. They are the essential underlying characteristics of the *c*-axis transport for all HTCSs.

## 1. Introduction

Planar doped-hole concentration ($P_{pl}$) is one of the most important physical parameters that dictate both normal and superconducting properties of high-temperature cuprate superconductors (HTCS). In order to systematically compare the doping dependence of various in-plane properties among different HTCSs, we have constructed the hole-scale based on the universal relation between the *in-plane* thermoelectric power at 290 K ($S^{290}$) and $P_{pl}$ [1]. The hole-scale is derived from, therefore consistent with, the relation between $S^{290}$ and $P_{pl}$ of single-layer $La_{2-x}Sr_xCuO_4$ ($P_{pl} = x$) and double-layer $Y_{1-x}Ca_xBa_2Cu_3O_6$ ($P_{pl} = x/2$). Until now, although there have been many temperature ($T$)-dependent studies of $c$-axis dc conductivity ($\sigma_c$) [2-38], there are few qualitatively doping-dependent studies of $\sigma_c$ for many different HTCSs [39, 40]. In the $T$-dependent studies one has to deal with many $T$-dependent factors involving possible changes of electronic band structure, of scattering rate and structural phase transitions. On the other hand, the doping-dependent studies have the distinct advantages of directly studying the evolution of the electronic states upon hole-doping without the complications due to the variation of $T$. Doping induced properties that are generic to all materials will be the real intrinsic properties of HTCS which warrant serious considerations.

In this paper, we have analyzed $\sigma_c$ of 19 hole-doped HTCSs from 37 published papers [2-38]. We found that the doping dependence of $\sigma_c$ for HTCSs is well scaled by $P_{pl}^{opt}$, where the HTCS material has the highest $T_c$, and $\sigma_c^{opt}$, which is the conductivity at $P_{pl}^{opt}$. The universal scaling behavior suggests that the intrinsic contributions of the doping-dependent-activated gap, the exponential doping dependences and the tunneling between adjacent $CuO_2$ block layers are essential and underlying the $c$-axis transport for all HTCSs.

## 2. Analysis method

Based on the hole-scale proposed in [1], we use two methods to extract the doped-hole concentration from the published data: $P_{pl}$ is determined from the value of $S^{290}$ by using the formula reported in [1] and, as the first and preferred method, $P_{pl}$ so determined is most reliable. Using first method we found that the $T_c$ versus $P_{pl}$ for HTCSs can have three types; a well-known symmetric dome-shaped $T_c$-$P_{pl}$ curve for $La_{2-x}Sr_xCuO_4$, a double-plateau $T_c$- $P_{pl}$ curve for $YBa_2Cu_3O_{6+\delta}$, and an asymmetric half-dome-shaped $T_c$-$P_{pl}$ curve for the other HTCSs [41]. We use the $T_c$-$P_{pl}$ curve as a secondary standard, the second method, and $P_{pl}$ can be determined from $T_c$ by comparing it with the corresponding $T_c$-$P_{pl}$ curve.

## 3. Results and discussion

In Fig. 1(a) we present $\sigma_c$ at 290 K ($\sigma_c(290)$) versus $P_{pl}/P_{pl}^{opt}$ for major single-, double- and triple-layer HTCSs on a semi-logarithmic plot. The $\sigma_c(290)$ determined in the optical conductivity are also plotted as star symbols [13,19]. It can be clearly seen that $\sigma_c(290)$ for underdoped HTCS are well described by five parallel lines from line I to line V. The double-layer $YBa_2Cu_3O_{6+\delta}$ and single-layer $La_{2-x}M_xCuO_4$ (M = Sr, Ca and Ba) are located on the line I and line II, respectively. The relatively few available data points of the Hg-based family are located around line III. The single-layer $Bi_2(Sr_{2-x}La_x)CuO_{6+\delta}$ and double-layer $Bi_2Sr_2Ca_{1-x}Gd_xCu_2O_{8+\delta}$ lie on the line IV together with the single-layer $Tl_2Ba_2CuO_{6+\delta}$ and $Ca_{2-x}Na_xCuO_2Cl_2$. On the line V, we can find the double-layer $Bi_2Sr_2CaCu_2O_{8+\delta}$ and triple-layer $Bi_2Sr_2Ca_2Cu_3O_{10+\delta}$. We would like to emphasize that since $P_{pl}^{opt}$ depends on the materials, which vary from 0.16 to 0.28 [41], therefore there is no such systematic behavior in the plot of $\sigma_c$ versus $P_{pl}$. We have previously reported that *c*-axis penetration depth for the single-layer HTCS is well correlated with, not $P_{pl}$, $P_{3D}$ [43]. $P_{3D}$ is an effective three dimensional (3D) doped-hole concentration defined as $P_{3D} = n_{lay}P_{pl}/V_c$ where $n_{lay}$ and $V_c$ are number of $CuO_2$ layer per unit cell and unit cell volume, respectively. Similarly, as shown in the plot of $\sigma_c(290)$ versus $P_{3D}$ in the inset, single-layer HTCSs lie on the three parallel red lines, although the multi-layer HTCSs also lie on another parallel black lines. They do not have same slopes as that in Fig. 1(a). Accordingly, the universal parallel behavior is ***only*** observable by using $P_{pl}/P_{pl}^{opt}$. Here, we define a unified doped-hole concentration ($p_u$) as $p_u \equiv P_{pl}/P_{pl}^{opt}$ [41].

In Fig. 1(b) we shows $\sigma_c(290)/\sigma_c^{opt}(290)$ as a function of $p_u$ on a semi-logarithmic plot. All $\sigma_c/\sigma_c^{opt}$'s on the $p_u$ range from ~0.2 to ~1.1 are seen to lie on one common linear relation. Although the scaling behavior was observed for 50 K $\leq T \leq$ 400 K, the slope ($\alpha = d[ln(\sigma_c/\sigma_c^{opt})]/dp_u$) increased with reducing $T$. The result at 110 K is shown in the inset. For $p_u <$ ~0.2, $\sigma_c$ rapidly deviates downward from the linear relation. For now, we will only focus on the universal scaling behavior of $\sigma_c$ in the underdoped HTCS. From Fig. 1(b), the *c*-axis dc conductivity $\sigma_c(p_u,T)$ can be expressed in terms of $\alpha(T)$ and $p_u$ as

$$\sigma_c(p_u,T)/\sigma_c^{opt}(T) = exp\{\alpha(T)(p_u-1)\} \qquad (0.2 < p_u < 1.1) \qquad (1)$$

It is indeed very surprisingly, in light of the many different types of block layers and crystal structures involved, $\sigma_c(p_u,T)$ for almost all major HTCSs can be described by formula (1).

It is worthy to point out that the effect of Zn doping on $\sigma_c$ is very small in the $YBa_2Cu_3O_{6+\delta}$ [44]. The Zn-dopant selectively occupies the Cu site in the Cu-$O_2$ plane. Since the doped-hole concentration does not depend on the Zn-doping, we conclude that the observed scaling behavior is derived only from the doped holes in the $CuO_2$ plane and it is *independent* of the *in-plane* disorder. Furthermore, since cation dopants simultaneously induce *out-of-plane* disorder with hole-doping, therefore, we conclude that the present scaling behavior is *independent* of the *out-of-plane* disorder also.

In Fig. 2 we plot $\alpha$ as a function of $1/T$. This plot manifests the tendency that the $\sigma_c$ changes from the thermally activated behavior at high-$T$ into the almost $T$-independent behavior at the low-$T$. For $T > 180$ K we have $\alpha(T) = 3.8 + E_g/T$. The activation energy gap $E_g$ is 460 K ~ 40 meV. For $T < 180$ K, $\alpha(T)$ deviates from the activation behavior with reducing $T$ and approaches to a constant value of 7.3 for $T \leq 70$ K. This cross-over from a activated behavior to a $T$-independent $\alpha$ ($\alpha \sim 7.3$) can be clearly seen in the linear plot in the inset of Fig. 2. A constant $\alpha$ implies that $\sigma_c$ has identical $T$-dependence as $\sigma_c^{opt}$ for $T \leq 70$ K.

In Fig. 3(a), we plot $\sigma_c^{opt}$, which is used as a scaling factor, as a function of the $CuO_2$ block distance ($d_{CC}$). Here, we define $d_{CC}$ to be copper to copper distance between adjacent $CuO_2$ block layers, consisting of single or multiple $CuO_2$ planes, sandwiched by “spacing layer”. The spacing layer is the insulating tunneling layer between two $CuO_2$ block layers [45]. To define $d_{CC}$, we use the crystal structures and the lattice constants compiled in [45-48]. For the double-layer HTCS, we define the $CuO_2$ block layer as the up-side pyramidal plus down-side pyramidal $CuO_2$ layers. But, for $YBa_2Cu_3O_{6+\delta}$ with CuO chain plane, since the scaling behavior of $\sigma_c$ is valid for $0.2 < p_u < 1.1$ irrespective of the formation of the chain ordering, we define $d_{CC} = 3.4$ Å which is the distance between two $CuO_2$ block layers consisting of two pyramidal $CuO_2$ planes and CuO chain plane that are separated by Y ion. For $d_{CC}$ varying from 3.4 Å of $YBa_2Cu_3O_{6+\delta}$ till 15.6 Å of $Bi_2Sr_2CaCu_2O_{8+\delta}I_x$, $\sigma_c^{opt}(290)$ and $\sigma_c^{opt}(110)$ on a semi-logarismic plot linearly decreases with increasing $d_{CC}$. However, $\sigma_c^{opt}$ of $Bi_2Sr_2CaCu_2O_{8+\delta}$, $Bi_2Sr_2Ca_2Cu_3O_{10+\delta}$ and $Ca_{2-x}Na_xCuO_2Cl_2$ are located out of the straight line in Fig. 3(a).

In the inset of Fig. 3(b) we plot $\sigma_c^{opt}$ as a function of $T$. Although $\sigma_c^{opt}(T)$ for the line I, II and IV show a small $T$-dependence below 110 K, practically, we can ignore the $T$-dependence for 110 K < $T \leq$ 400 K. Accordingly, $\sigma_c^{opt}$ is approximately represented by the following $T$-independent formula,

$$\sigma_c^{opt}(T) = 1479\ exp(-0.6 d_{CC}) \quad [\Omega^{-1}cm^{-1}] \qquad (110\ K < T \leq 400\ K) \qquad (2)$$

Noted that the formula (2) strongly suggests that there is a universal intrinsic contribution of $T$-independent quantum tunneling to $\sigma_c$ for 110 K < $T \leq$ 400 K. For instance, inter-layer tunneling through a square potential barrier with barrier height $h$ and barrier width $d$ will have a conductivity $\sigma \sim exp(-hd)$ [49]. Therefore we can consider 0.6 to be an effective universal intrinsic barrier height for almost all underdoped HTCSs. This universal barrier height implies a universal kinetic energy of the tunneling particles that contributes to $\sigma_c$. Accordingly, the term of $exp(-0.6d_{CC})$ can be interpreted as the inter-layer tunneling between the conducting $CuO_2$ block layers through the spacing layer along $c$-axis. In Fig. 3(a), the $\sigma_c^{opt}$'s of $Bi_2Sr_2CaCu_2O_{8+\delta}$, $Bi_2Sr_2Ca_2Cu_3O_{10+\delta}$ and $Ca_{2-x}Na_xCuO_2Cl_2$ seem to be anomalous. However if we define the $CuO_2$ block layer as the up-side (or down-side) pyramidal $CuO_2$ layer plus the square $CuO_2$ layer for $Bi_2Sr_2Ca_2Cu_3O_{10+\delta}$ and one up-side (or down-side) $CuO_2$ layer for $Bi_2Sr_2CaCu_2O_{8+\delta}$, then $d_{CC}$ = 14.7 Å for both materials. In Fig. 3(c), we can see that both Bi-based systems fall back on the common straight line. We do not know why $Ca_{2-x}Na_xCuO_2Cl_2$ deviates from the universal straight line. It may be due to the electronic effect of Cl ion in tunneling matrix since all apical oxygen on the $CuO_2$ planes are substituted by Cl ion. Now it becomes clear that the parallel downshift of the conductivity from line I to line V in Fig. 1 are exclusively structural in origin. They come from the contribution of the tunneling term due to the increase of $d_{CC}$ from line I to line V.

If we plug formula (2) into (1) by using the relation of $\alpha(T) = 3.8 + E_g/T$ ($T$ > 180 K) we arrive at the following final universal expression of $\sigma_c(p_u,T)$ $[\Omega^{-1}cm^{-1}]$ for almost all HTCSs.

$$\sigma_c(p_u,T) = \begin{cases} 1479\exp\{-3.8(1-p_u)\}\exp\{-E_g(1-p_u)/T\}\exp(-0.6d_{CC}) & (180K < T \le 400K) \\ 1479\exp\{-\alpha(T)(1-p_u)\}\exp(-0.6d_{CC}) & (110K < T \le 180K) \\ \sigma_c^{opt}(T)\exp\{-\alpha(T)(1-p_u)\} & (70K < T \le 110K) \\ \sigma_c^{opt}(T)\exp\{-7.3(1-p_u)\} & (T \le 70K) \end{cases} \quad (3)$$

For $T$ > 180 K, $\sigma_c$ approaches $\sigma_c^{opt}$ through contributions from a linearly decreasing doping-dependent activated gap $E_g(1-p_u)$ and an exponentially increasing term, $exp\{-3.8(1-p_u)\}$, with increasing doping. If we fix $p_u$, then $\sigma_c$ decreases exponentially with decreasing $T$ as observed universally in $\sigma_c$ of underdoped HTCSs. Irrespective of $p_u$, for $T$ > 110 K, $\sigma_c$ decreases exponentially with increasing $d_{CC}$. Activated gap, vanishing at $p_u$ = 1, is consistent with the observation of the metallic $\sigma_c$ in many optimally doped HTCSs.

The activation and quantum tunneling terms in formula (3) have interesting implications, namely, it suggests a universal intrinsic quantum tunneling transport mechanism along *c*-axis. Indeed, in a recent intrinsic interlayer (*c*-axis) tunneling experiment of $Bi_2Sr_2CaCu_2O_{8+\delta}$ crystal clearly showed a current-voltage characteristics change from a thermal activated like behavior above $T_c$ to a $T$-independent quantum tunneling like behavior below $T_c$ in the underdoped regime [50]. The quantum tunneling like contribution along *c*-axis was attributed to the tunneling that involved cooper pairs below $T_c$. This result can be easily understood in terms of our universal *c*-axis transport since in formula (3) we see that, for $T$ > 180 K, the thermally activated carriers tunnel through the insulating spacing layer defined for Fig. 3. Upon lowering temperature below 180 K the contribution of $T$-independent tunneling starts to win over the activated contribution so that $\sigma_c$ approaches the final $T$-independent tunneling $\sigma_c(p_u,T) = 1479exp\{-7.3(1-p_u)\}exp(-0.6d_{CC})$ around 110 K. Below 110 K, from the formula (3), the $T$-dependence of $\sigma_c$ is eventually dominated by that of $\sigma_c^{opt}$. The tunneling barrier is obviously different from the activated gap since it is doping independent and does not vanish as the activated gap at $p_u$ = 1. Therefore we conclude that there is already $T$-independent quantum tunneling existing in the *c*-axis transport even at 180 K.

## 4. Conclusion

We have extracted, using $p_u$ and the $\sigma_c$ data accumulated in the past twenty years, the universal scaling behavior of *c*-axis dc conductivity $\sigma_c(p_u,T)/\sigma_c^{opt}(T) = exp\{\alpha(T)(p_u-1)\}$ for $0.2 < p_u < 1.1$. Without losing essential physics, we arrive at an universal expression of $\sigma_c = 1479 \exp\{\alpha(T)(p_u-1)-0.6d_{CC}\}$ [$\Omega^{-1}cm^{-1}$] where $\alpha(T) = 3.8 + E_g/T$ with an activation energy gap $E_g$ = 460 K for $T$ > 180 K and constant $\alpha$ = 7.3 for $T \leq$ 70 K. This suggests that the doping-dependent activated gap, exponential increase of $\sigma_c$ upon increasing $p_u$ and the exponential decrease of $\sigma_c$ upon increasing $d_{CC}$ are essential underlying characteristics of the *c*-axis transport for all HTCSs. Our observations indicate that there is a unified mechanism for *c*-axis transport. A coherent physical picture of understanding all these behaviors posts another challenge to the understanding of the electronic properties of layered cuprates and, possibly, holds the key to the mechanism of high-$T_c$.

**Figure caption:**

**Fig. 1** (a) $\sigma_c(290)$ versus $P_{pl}/P_{pl}^{opt}$ on a semi-logarithmic plot for major single, double and triple-layer HTCSs. The plotted data are sumarized in Table 1. The inset shows the plot of $\sigma_c(290)$ versus $P_{3D}$. (b) $\sigma_c(290)/\sigma_c^{opt}(290)$ versus $p_u$. The inset shows the plot of $\sigma_c(110)/\sigma_c^{opt}(110)$ versus $p_u$. The plotted data are the same as Fig. 1(a), excluding the materials with $T_c$ over 110 K.

**Fig. 2** $\alpha$ versus $1/T$. The inset shows $\alpha$ versus $T$. The solid line is calculated by a formula of $\alpha(T) = 3.8 + 460/T$.

**Fig. 3** (a) $\sigma_c^{opt}$ as a function of the $CuO_2$-block distances ($d_{CC}$). (b) $\sigma_c^{opt}$, which is used as a scaling factor, in log-scale as a function of $T$. (c) $\sigma_c^{opt}$ as a function of the new $d_{CC}$. The straight lines in (a) and (c) represents the formula (2). The straight lines in (b) are guide to the eyes

**Table caption:**

**Table 1.** The information for data plotted in Fig. 1.

**Table caption:**

**Table 1.** The information for data plotted in Fig. 1.

| line | chemical formula | definition of $P_{pl}$ | $P_{pl}^{opt}$ | Ref(s). |
|---|---|---|---|---|
| I | $YBa_2Cu_3O_{6+\delta}$ | $S^{290}$ | 0.25 | [10] |
| | | $T_c$ | 0.25 | [2-9,11-13] |
| | $Ca_{0.5}La_{1.25}Ba_{1.25}Cu_3O_y$ | $T_c$ | 0.235 | [14] |
| II | $La_{2-x}Sr_xCuO_4$ | Sr-content | 0.16 | [15-19] |
| | $La_{2-x}Ca_xCuO_4$ | Ca-content | 0.16 | [20] |
| | $La_{2-x}Ba_xCuO_4$ | Ba-content | 0.16 | [21,22] |
| | $La_{1.6-x}Nd_{0.4}Sr_xCuO_4$ | Sr-content | 0.16 | [23] |
| III | $HgBa_2CuO_{4+\delta}$ | $S^{290}$ | 0.235 | [19] |
| | $(Hg_{0.8}Cu_{0.2})Ba_2CuO_{4+\delta}$ | $S^{290}$ | 0.235 | [24,25] |
| | $HgBa_2Ca_2Cu_3O_{8+\delta}$ | $S^{290}$ | 0.215 | [26] |
| IV | $Tl_2Ba_2CuO_{6+\delta}$ | $T_c$ | 0.25 | [27] |
| | $Bi_2(Sr_{2-x}La_x)CuO_{6+\delta}$ | $T_c$ | 0.28 | [29] |
| | $Bi_2Sr_2Ca_{1-x}Gd_xCu_2O_{8+\delta}$ | $S^{290}$ | 0.25 | [30] |
| | $Br_2Sr_2Ca_{1-x}Er_xCu_2O_{8+\delta}$ | $T_c$ | 0.238 | [31] |
| | $Ca_{1-x}Na_xCuO_2Cl_2$ | Na-content | 0.19 | [32] |
| V | $Bi_2Sr_2CaCu_2O_{8+\delta}$ | $S^{290}$ | 0.238 | [33] |
| | | $T_c$ | 0.238 | [11,34-35] |
| | $Bi_2Sr_2Ca_2Cu_3O_{10+\delta}$ | $S^{290}$ | 0.215 | [36] |
| | | $T_c$ | 0.215 | [37] |
| | $(Bi_{2-x}Pb_x)Sr_2CaCu_2O_{8+\delta}$ | $T_c$ | 0.238 | [38] |
| | $(Bi_{2-x}Pb_x)Sr_2ErCu_2O_{8+\delta}$ | $S^{290}$ | 0.238 | [39] |
| | $Bi_2Sr_2CaCu_2O_{8+\delta}I_x$ | $S^{290}$ | 0.238 | [33] |

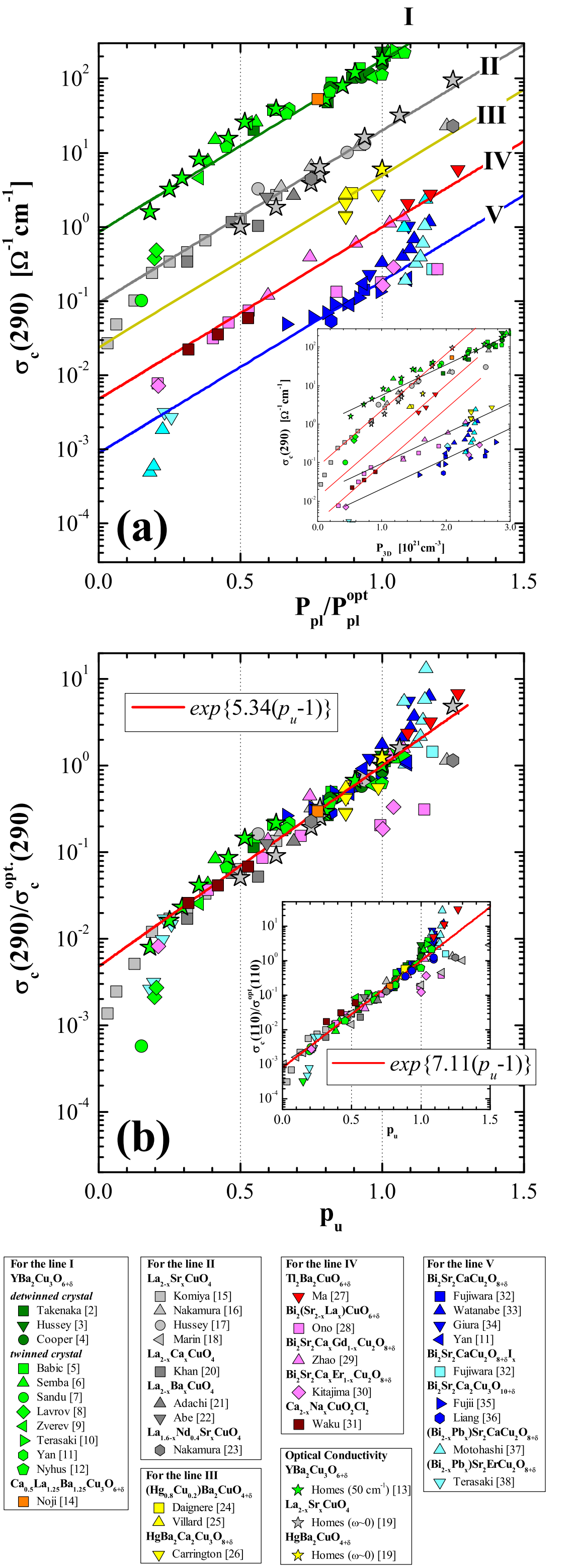

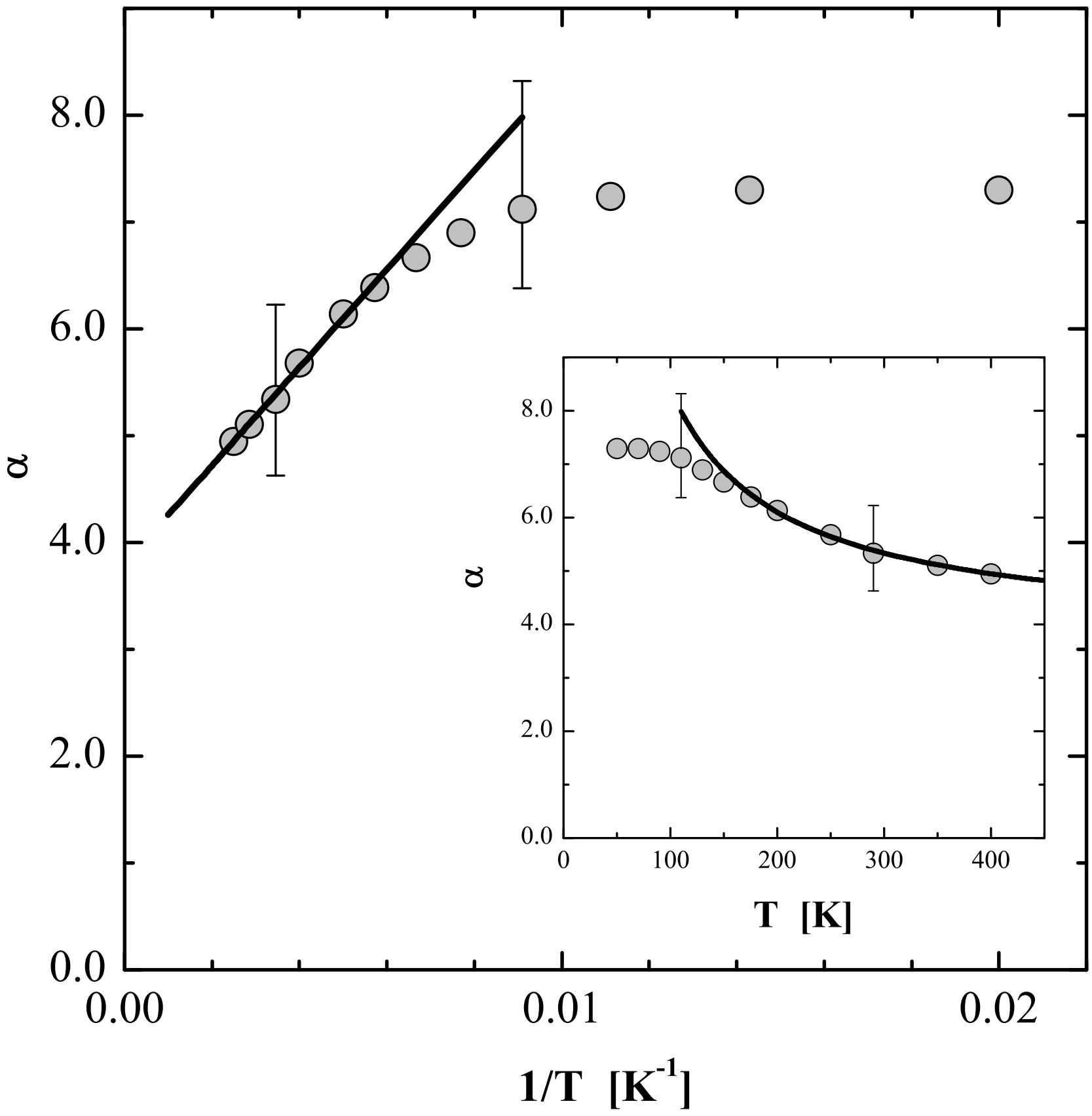

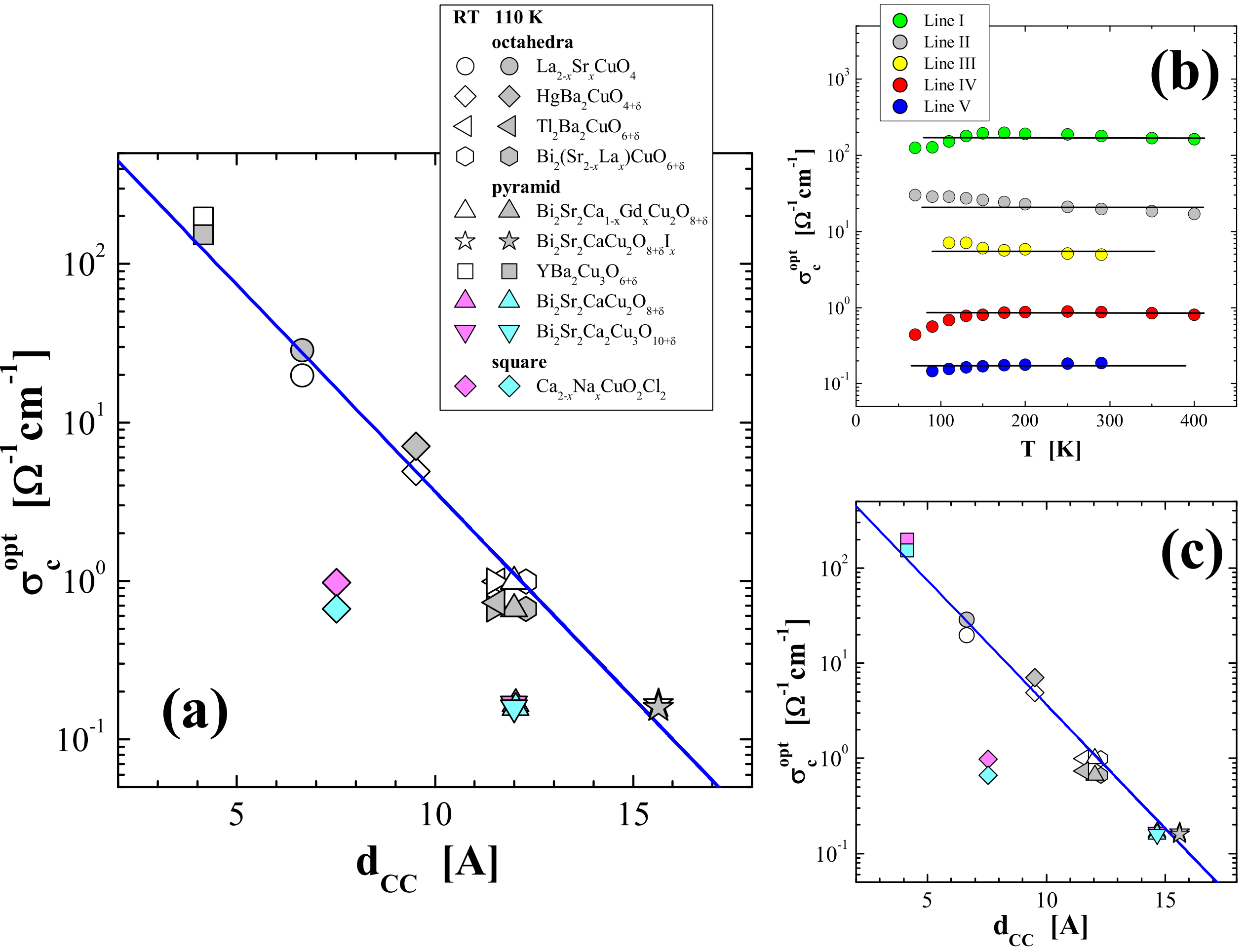